\documentclass[]{pasj01}

\Received{2020/03/06}
\Accepted{2020/05/13}
 
\begin{document} 

\title{ 
\LETTERLABEL 
%
Spatially Resolved Molecular Gas Properties of Host Galaxy of Type I Superluminous Supernova SN~2017egm
}

\author{Bunyo \textsc{Hatsukade}\altaffilmark{1}}
\email{hatsukade@ioa.s.u-tokyo.ac.jp}

\author{Kana \textsc{Morokuma-Matsui}\altaffilmark{1}}
\author{Masao \textsc{Hayashi}\altaffilmark{2}}
\author{Nozomu \textsc{Tominaga}\altaffilmark{3,4}}

\author{Yoichi \textsc{Tamura}\altaffilmark{5}}
\author{Kotaro \textsc{Niinuma}\altaffilmark{6}}
\author{Kazuhiro \textsc{Motogi}\altaffilmark{6}}
\author{Tomoki \textsc{Morokuma}\altaffilmark{1}}
\author{Yuichi \textsc{Matsuda}\altaffilmark{2,7}}

\altaffiltext{1}{Institute of Astronomy, Graduate School of Science, The University of Tokyo, 2-21-1 Osawa, Mitaka, Tokyo 181-0015, Japan}
\altaffiltext{2}{National Astronomical Observatory of Japan, 2-21-1 Osawa, Mitaka, Tokyo 181-8588, Japan}
\altaffiltext{3}{Department of Physics, Faculty of Science and Engineering, Konan University, 8-9-1 Okamoto, Kobe, Hyogo 658-8501, Japan}
\altaffiltext{4}{Kavli Institute for the Physics and Mathematics of the Universe (WPI), The University of Tokyo, 5-1-5 Kashiwanoha, Kashiwa, Chiba 277-8583, Japan}
\altaffiltext{5}{Department of Physics, Nagoya University, Furo-cho, Chikusa-ku, Nagoya 464-8602, Japan}
\altaffiltext{6}{Graduate School of Science and Engineering, Yamaguchi University, Yoshida 1677-1, Yamaguchi, Yamaguchi 753-8512, Japan}
\altaffiltext{7}{Graduate University for Advanced Studies (SOKENDAI), Osawa 2-21-1, Mitaka, Tokyo 181-8588, Japan}

\maketitle

\begin{abstract}
We present the results of CO(1--0) observations of the host galaxy of a Type I superluminous supernova (SLSN-I), SN~2017egm, one of the closest SLSNe-I at $z = 0.03063$, by using the Atacama Large Millimeter/submillimeter Array. 
The molecular gas mass of the host galaxy is $M_{\rm gas} = (4.8 \pm 0.3) \times 10^9$~$M_{\odot}$, placing it on the sequence of normal star-forming galaxies in an $M_{\rm gas}$--star-formation rate (SFR) plane. 
The molecular hydrogen column density at the location of SN~2017egm is higher than that of the Type II SN PTF10bgl, which is also located in the same host galaxy, and those of other Type II and Ia SNe located in different galaxies, suggesting that SLSNe-I have a preference for a dense molecular gas environment. 
On the other hand, the column density at the location of SN~2017egm is comparable to those of Type Ibc SNe. 
The surface densities of molecular gas and the SFR at the location of SN~2017egm are consistent with those of spatially resolved local star-forming galaxies and follow the Schmidt--Kennicutt relation. 
These facts suggest that SLSNe-I can occur in environments with the same star-formation mechanism as in normal star-forming galaxies. 
\end{abstract}

\begin{figure*}
\begin{center}
\includegraphics[width=.742\linewidth]{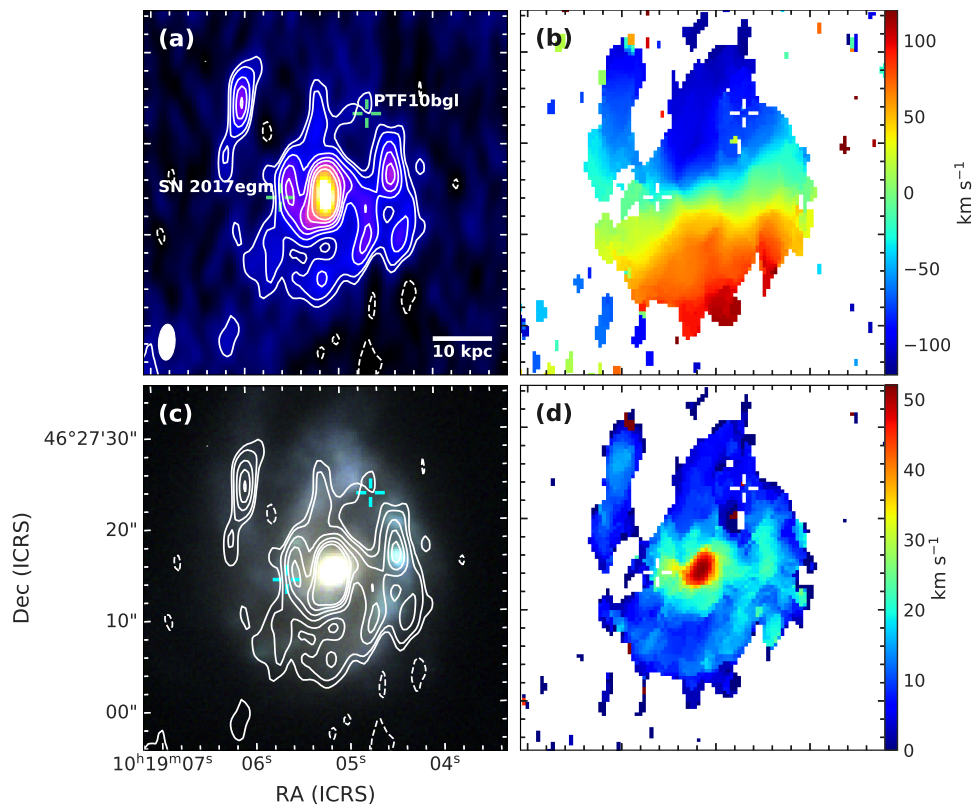}
\end{center}
\caption{
(a) CO(1--0) velocity-integrated intensity map. 
The contours are $-3, -2, 2, 3, 5, 7, 9, 12$, and $15\sigma$. 
Hair marks represent the positions of SN~2017egm and PTF10bgl. 
The synthesized beam size is shown in the lower-left corner. 
(b), (d) Maps of intensity-weighted velocity field and velocity dispersion. 
The emissions with $<$$3\sigma$ are clipped for the maps of velocity field and dispersion. 
(c) Pan-STARRS {\sl gri} color image. 
The contours are the same as in (a). 
}
\label{fig:map}
\end{figure*}

\section{Introduction}

Superluminous supernovae (SLSNe) are extremely luminous explosions with absolute magnitudes of $\lesssim$$-21$~mag, which are $\sim$10--100 times brighter than typical Type Ia and core-collapse SNe \citep{gal12}. 
SLSNe are a new class of SNe that was discovered only recently by wide-field, untargeted, time-domain surveys (e.g., \cite{quim07, quim11}). 
They are detected from local ($z = 0.03$) to high-redshift galaxies ($z \sim 4$; \cite{cook12}), and therefore can be powerful indicators of environments in the distant universe. 
SLSNe are classified into two main subclasses depending on the presence of hydrogen signatures in the observed spectra: hydrogen-poor Type I (SLSN-I) and hydrogen-rich Type II (SLSN-II) \citep{gal12}.
Due to their huge luminosity and scarcity, the physical nature of SLSNe is still a matter of debate, and especially SLSNe-I are among the least understood SN populations.

Spatially resolving observations of molecular gas provide the physical properties of the interstellar medium (ISM) in the local environment of stellar explosions, such as molecular gas content, star-formation efficiency, and velocity field (e.g., \cite{galb17, arab19, moro19}). 
\citet{arab19} conducted CO(1--0) observations of the host galaxy of a SLSN-II, PTF10tpz, at $z = 0.03994$ with the Atacama Large Millimeter/submillimeter Array (ALMA), and found that PTF10tpz is located close to the intersection of the gas lanes and the inner structure of the host galaxy. 
They suggested that in situ formation of massive stars due to the internal dynamics of the host galaxy and high densities are favorable conditions for the formation of SLSN progenitors.

SN~2017egm/Gaia17biu at $z = 0.03063$, one of the closest SLSNe-I, was discovered on May 23, 2017 \citep{dong17, sdss17}. 
The host galaxy, NGC~3191, is a massive spiral galaxy ($M_* = 5 \times 10^{10}$ $M_{\odot}$) with active star formation (SFR $\sim 5$--15 $M_{\odot}$~yr$^{-1}$) \citep{stol13, nich17, chen17a, bose18}. 
The metallicity at the SN site shows a (super-)solar metallicity ($\sim$1.3--2.6 $Z_{\odot}$; \cite{nich17, chen17a, bose18}), while there is a work showing a sub-solar metallicity (0.6 $Z_{\odot}$; \cite{izzo18}). 
It is notable that NGC~3191 also hosted two other SNe: SN~1988B (Type Ia) and PTF10bgl (Type II). 
SN 1988B was reported to be located at $10''$ north of the galaxy center \citep{schi88, fill88}, although the precise location was not provided. 
PTF10bgl was located $\sim$10$''$ north-west of the galaxy center \citep{arca10}. 
This enables us to compare the environments between a SLSN-I and a Type II SN located in the same galaxy.

In this Letter, we present the results of ALMA CO(1--0) observations of the host galaxy of SN~2017egm. 
This is the first study on molecular gas in a SLSN-I host galaxy. 
Throughout the paper, we adopt the cosmological parameters $H_0=67.8$ km s$^{-1}$ Mpc$^{-1}$, $\Omega_{\rm{M}}=0.308$, and $\Omega_{\Lambda}=0.692$ \citep{plan16}. 
The luminosity distance to the host galaxy is 138.7 Mpc, and $1''$ corresponds to 0.65 kpc.

\section{Observations and Results} \label{sec:observations}

ALMA CO(1--0) observations were conducted on Mar. 28 and 29, 2019, for a Cycle 6 program (Project code: 2018.1.00370.S). 
The redshifted CO(1--0) line was observed with Band 6. 
The correlator was used in the time domain mode with a bandwidth of 1875 MHz (488.28~kHz $\times$ 3840 channels). 
Four basebands were used, providing a total bandwidth of 7.5 GHz. 
The array configuration was C43-2 with baseline lengths of 15.0--457.3 m. 
The number of available antenna was 46--48, and the on-source integration time was 79 min. 
Bandpass and flux calibrations were performed with J1058+0133 and phase calibrations with J0927+3902.

The data were reduced with Common Astronomy Software Applications (CASA; \cite{mcmu07}). 
Maps were processed with a \verb|tclean| task with \verb|Briggs| weighting and a \verb|robust| parameter of 0.5. 
The synthesized beamsize is $3\farcs9 \times 1\farcs8$ (2.6 kpc $\times$ 1.2 kpc) with a position angle of $-3.7^{\circ}$. 
The rms noise level is 1.5 mJy~beam$^{-1}$ for a spectrum with a velocity resolution of 5 km~s$^{-1}$.

Figure~\ref{fig:map} shows the obtained maps of CO(1--0) velocity-integrated intensity, intensity-weighted velocity field, and intensity-weighted velocity dispersion. 
The CO emission is clearly detected with a smooth rotation signature, which is consistent with the H$\alpha$ IFU observations \citep{chen17a}. 
The bright CO peak $\sim$$7''$ west of the galaxy center coincident with an {\sc Hii} region \citep{chen17a, izzo18} and the brightest peak of a 10 GHz continuum map \citep{bose18}. 
SN~2017egm is located close to a bright CO blob east of the galaxy center. 
The CO emission is also detected at the location of PTF10bgl at the $\sim$$2\sigma$ level. 
\citet{izzo18} found a tangential or warp-like disturbance, based on a detailed kinemetric analysis on the H$\alpha$ map, and suggest that this could be a sign of interaction with its companion, MCG$+$08-19-017, at a projected distance of $\sim$45 kpc and a radial velocity difference of $\sim$200 km s$^{-1}$. 
We do not find any atypical feature in the CO maps around the location of SN~2017egm or PTF10bgl.

\section{Discussion} \label{sec:discussion}

\subsection{Host Galaxy} \label{sec:host}

\begin{figure}
\begin{center}
\includegraphics[width=.92\linewidth]{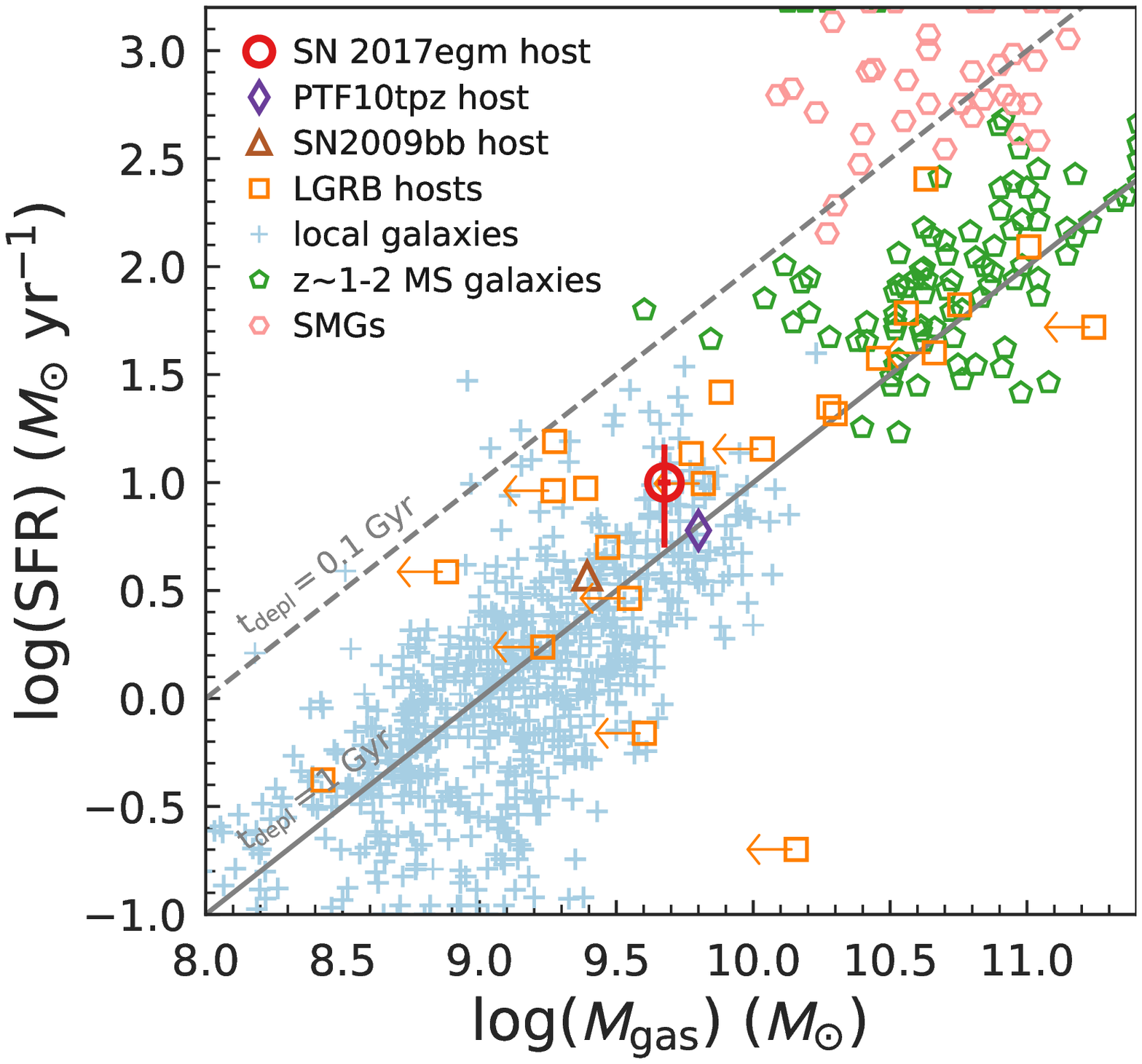}
\end{center}
\caption{
Comparison of molecular gas mass and SFR. 
The vertical bar for the SN~2017egm host shows the range of SFR in the literature, 
while the horizontal bar shows the error caused by flux measurement uncertainty (1$\sigma$). 
For comparison, we plot 
the PTF10tpz (SLSN-II) host galaxy \citep{arab19}, 
the SN~2009bb (broad-line Ic) host galaxy \citep{mich18}, 
the host galaxies of long-duration GRBs (arrows are upper limits) compiled by \citet{hats20}, 
local galaxies \citep{sain11, sain17, both14}, 
$z \sim 1$--2 main-sequence galaxies \citep{tacc13, seko16}, 
and submillimeter galaxies \citep{both13}. 
The solid and dashed lines represent gas depletion times of 0.1 and 1 Gyr, respectively. 
}
\label{fig:mgas-sfr}
\end{figure}

The CO luminosity of the host galaxy is calculated to be $L'_{\rm CO} = (1.1 \pm 0.1) \times 10^9$~$L_{\odot}$ following the equation of \citet{solo05}. 
The molecular gas mass is $M_{\rm gas} = (4.8 \pm 0.3) \times 10^9$~$M_{\odot}$ derived from $M_{\rm gas} = \alpha_{\rm CO} L'_{\rm CO}$, 
where $\alpha_{\rm CO}$ is a CO-to-H$_2$ conversion factor including the contribution of the helium mass. 
The conversion factor can vary with different environments (see, e.g., \cite{bola13} for a review). 
The conversion factor is thought to be dependent on gas-phase metallicity, increasing $\alpha_{\rm CO}$ with decreasing metallicity (e.g., \cite{wils95, bola13}).
Because the host galaxy has a metallicity close to the solar value, we adopt a Galactic conversion factor of $\alpha_{\rm CO} = 4.3$ $M_{\odot}$~(K~km~s$^{-1}$~pc$^2$)$^{-1}$ (with 30\% uncertainty; \cite{bola13}). 
The derived physical quantities are presented in Table~\ref{tab:results}. 
Note that errors take into account only flux measurement uncertainties. 
The molecular gas mass fraction ($\mu_{\rm gas} = M_{\rm gas}/M_*$) is 0.095, which is comparable to those of local star-forming galaxies with a similar stellar mass \citep{sain11, sain17, both14}. 
The molecular gas mass is compared with the SFR in Figure \ref{fig:mgas-sfr}. 
Because the SFR of the host galaxy ranges from 5 to 15 $M_{\odot}$~yr$^{-1}$ in the literature \citep{stol13, nich17, chen17a}, we adopt the range as a vertical line in the plot.  
The host galaxy is located in a similar region for local galaxies and on the sequence of normal star-forming galaxies. 
The gas depletion timescale ($\tau_{\rm gas} = M_{\rm gas}$/SFR) is 0.32--0.95 Gyr, which is comparable to those of local star-forming galaxies with a similar stellar mass \citep{both14, sain17}. 
The gas depletion timescale is also comparable to the host galaxies of PTF10tpz (SLSN-II; \cite{arab19}) and SN~2009bb (broad-line Ic SN; \cite{mich18}).

\begin{table}
\tbl{Derived properties of the host galaxy and at the sites of SN~2017egm and PTF10bgl}{
\begin{tabular}{lll}
\hline
Host galaxy & $L'_{\rm CO}$ (K km s$^{-1}$~pc$^2$) & $(1.1 \pm 0.1) \times 10^9$ \\
            & $M_{\rm gas}$ ($M_{\odot}$) & $(4.8 \pm 0.3) \times 10^9$ \\
            & $\mu_{\rm gas}$\footnotemark[$*$] & 0.095 \\
            & $\tau_{\rm depl}$\footnotemark[$\dag$] (Gyr) & 0.32--0.95 \\
            & SFE\footnotemark[$\dag$] (Gyr$^{-1}$) & 1.0--3.1 \\
\hline
SN~2017egm site & $N({\rm H_2})$ (cm$^{-2}$) & $(1.6 \pm 0.3) \times 10^{21}$ \\
                & $\Sigma_{\rm gas}$ ($M_{\odot}$~pc$^{-2}$) & $35 \pm 6$ \\
\hline
PTF10bgl site & $N({\rm H_2})$ (cm$^{-2}$) & $(5.6 \pm 2.7) \times 10^{20}$ \\
              & $\Sigma_{\rm gas}$ ($M_{\odot}$~pc$^{-2}$) & $12 \pm 6$ \\ 
\hline
\end{tabular}}\label{tab:results}
\begin{tabnote}
Errors take into account only flux measurement uncertainty (1$\sigma$). 
Galactic CO-to-H$_2$ conversion factor of $\alpha_{\rm CO}= 4.3$ $M_{\odot}$~(K~km~s$^{-1}$~pc$^2$)$^{-1}$ is assumed. \\
\footnotemark[$*$] Molecular gas fraction ($M_{\rm gas}/M_*$). \\
\footnotemark[$\dag$] Gas depletion timescale ($\mu_{\rm gas} = M_{\rm gas}$/SFR) and star-formation efficiency (SFE $=$ SFR/$M_{\rm gas}$) assuming SFR = 5--15 $M_{\odot}$~yr$^{-1}$ based on the measurements in previous studies. \\
\end{tabnote}
\end{table}

\subsection{SLSN Site} \label{sec:site}

\begin{figure}
\begin{center}
\includegraphics[width=.86\linewidth]{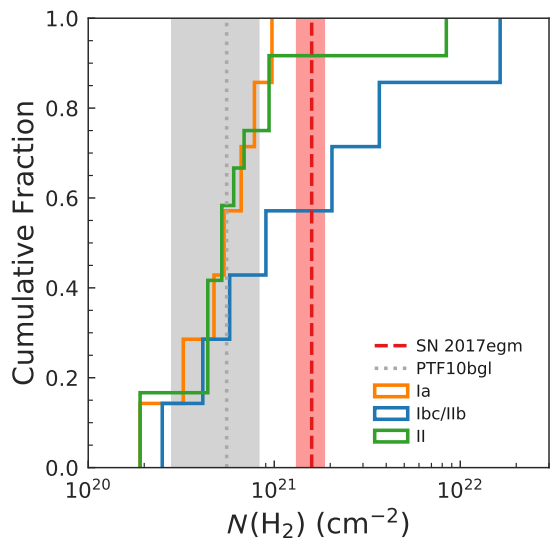}
\end{center}
\caption{
Cumulative distribution of molecular column density $N({\rm H_2})$ for the three SN types (Ia, Ibc/IIb, and II) including upper limits derived from spatially resolved observations of SN hosts by \citet{galb17}. 
Vertical lines represent the column densities at the positions of SN~2017egm and PTFbgl. 
Shaded regions show errors caused by flux measurement uncertainty (1$\sigma$). 
}
\label{fig:nH2}
\end{figure}

The metallicity at the SN~2017egm site measured in previous studies is controversial. 
\citet{nich17} and \citet{chen17a} showed a (super-)solar metallicity of $12+\log{\rm (O/H)} = 8.8$ and $9.11$, respectively, using the $R_{23}$ diagnostic with the \citet{kobu04} calibration. 
\citet{bose18} also found a super-solar metallicity of $12+\log{\rm (O/H)} = 9.0$ using [{\sc Nii}]/H$\alpha$ diagnostic with the \citet{naga06} calibration. 
On the other hand, \citet{izzo18} found a sub-solar metallicity of $12+\log{\rm (O/H)} = 8.49$ and 8.45 using the N2 and O3N2 diagnostics, respectively, based on the calibrations of \citet{mari13}. 
It is know that metallicity diagnostics are uncertain (e.g., \cite{kewl08}) and the differences in the previous studies can be due to different diagnostics \citep{chen17a, izzo18}. 
In oder to see the effect of metallicity on $\alpha_{\rm CO}$, we apply the relation between metallicity and $\alpha_{\rm CO}$ of \citet{genz15}, where they took the geometric mean of the empirical relations of \citet{genz12} and \citet{bola13} and derived the relation for the local and high-redshift sample. 
To apply the relation, we convert the metallicity to the calibration of \citet{pett04} by using the metallicity conversion of \citet{kewl08}.
The derived metallicity-dependent $\alpha_{\rm CO}$ is 3.4--6.6 $M_{\odot}$~(K~km~s$^{-1}$~pc$^2$)$^{-1}$. 
In the following discussions, we assume a Galactic $\alpha_{\rm CO}$ of 4.3 $M_{\odot}$~(K~km~s$^{-1}$~pc$^2$)$^{-1}$ (corresponding $X_{\rm CO}$ is $2 \times 10^{20}$ cm$^{-2}$ (K km s$^{-1}$)$^{-1}$), which is in the range of the metallicity-dependent conversion factor and is used in previous studies on the host galaxies of SNe \citep{galb17, mich18, arab19}\footnote{\citet{mich18} assumed a Galactic conversion factor of $\alpha_{\rm CO} = 5$ $M_{\odot}$~(K~km~s$^{-1}$~pc$^2$)$^{-1}$.}. 
The column densities of molecular gas at the positions of SN~2017egm and PTF10bgl are $N({\rm H_2}) = (1.6 \pm 0.3) \times 10^{21}$ cm$^{-2}$ and $(5.6 \pm 2.7) \times 10^{20}$ cm$^{-2}$, respectively. 
Here we adopt the same $\alpha_{\rm CO}$ for both the SN sites, because \citet{izzo18} found that metallicities at the sites are similar. 
We note that even if we assume the higher $\alpha_{\rm CO}$, the following discussions and conclusions would not change.
The column density at the SN~2017egm site is found to be higher than that of the PTF10bgl site by a factor of three. 
We compare the column densities with the results of spatially resolving CO(1--0) observations of host galaxies of Type Ia, Ibc/IIb, and II SNe in \citet{galb17}. 
Figure~\ref{fig:nH2} shows the cumulative distributions of $N({\rm H_2})$ for the SNe. 
The vertical lines represent the values for SN~2017egm and PTF10bgl obtained in this study. 
We find that the column density at the SN~2017egm site is higher than those of SNe Ia and II, suggesting that SLSN-I progenitors have a preference for a higher-molecular-gas-density environment. 
The higher surface density of molecular gas is also reported for PTF10tpz, a SLSN-II, by \citet{arab19}. 
This appears to suggest that a dense molecular gas environment is an important factor for producing SLSN progenitors.

\begin{figure}
\begin{center}
\includegraphics[width=.92\linewidth]{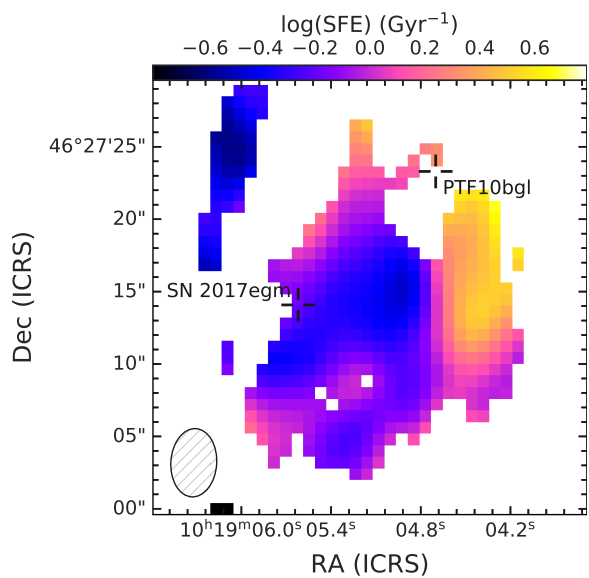}
\end{center}
\caption{
Map of star-formation efficiency (SFR/$M_{\rm gas}$) in the host galaxy derived from the molecular gas surface density map base on the CO(1--0) map and the SFR map based on the MaNGA H$\alpha$ observations \citep{chen17a}. 
The region where the CO(1--0) velocity-integrated intensity map is above $2\sigma$ is presented. 
The spatial resolution is shown in the lower-left corner. 
}
\label{fig:sfe}
\end{figure}

\begin{figure}
\begin{center}
\includegraphics[width=.92\linewidth]{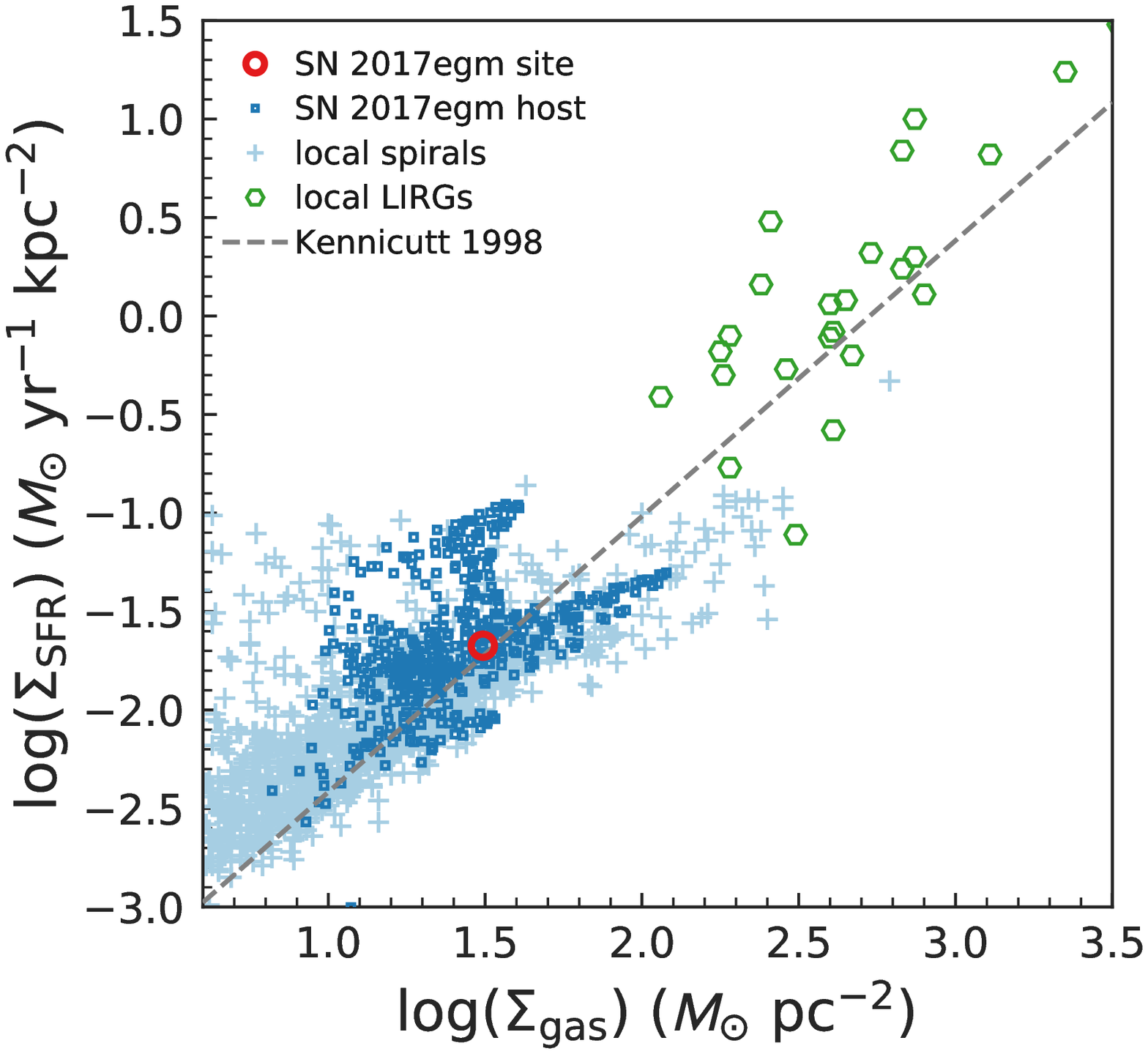}
\end{center}
\caption{
Comparison of molecular gas mass surface density and SFR surface density. 
The data points for the SN~2017egm host galaxy are measured at pixels where the CO(1--0) velocity-integrated intensity map is above $2\sigma$. 
For comparison, we plot other type of galaxies in the literature, where size measurements are available: 
local spirals \citep{kenn98a, bigi10}, 
and local LIRGs \citep{kenn98a}. 
The dashed line represents the relation of \citet{kenn98b}. 
}
\label{fig:sigmaGas-sigmaSFR}
\end{figure}

On the other hand, the column density at the SN~2017egm site is comparable to the median value of Type Ibc/IIb SNe ($N({\rm H_2}) =  1.5 \times 10^{21}$ cm$^{-2}$ for six CO-detected SN sites; \cite{galb17}). 
The molecular gas surface density is an order of magnitude lower than at the PTF10tpz site ($\Sigma_{\rm gas} \sim 700$ $M_{\odot}$~pc$^{-2}$ over $\sim$350 pc scale; \cite{arab19}), where the SLSN occurred near the intersection region of gas lanes and the inner structure in the host galaxy. 
Note that although the gas surface density at the PTF10tpz site is corrected for the inclination of the host galaxy \citep{arab19}, its large inclination angle of $68^{\circ}$ makes it difficult to estimate the actual column density. 
Figure~\ref{fig:sfe} shows the map of the star-formation efficiency (SFE $=$ SFR/$M_{\rm gas}$) in the host galaxy. 
The map is created from the molecular gas surface density map based on our CO(1--0) observations and the SFR map based on the MaNGA H$\alpha$ observations by \citet{chen17a}. 
Both the maps are convolved with the beam of the other map to match the spatial resolution. 
The SFE at the location of SN~2017egm does not appear to be special within the host galaxy. 
This is illustrated in Figure~\ref{fig:sigmaGas-sigmaSFR}, which compares the surface densities of the molecular gas and the SFR. 
The pixel-by-pixel variations within the host galaxy are plotted. 
We used the region where the CO(1--0) velocity-integrated intensity map is above $2\sigma$. 
We also compare the results of spatially resolved (kpc-scale) observations of local star-forming galaxies. 
The location of SN~2017egm in Figure~\ref{fig:sigmaGas-sigmaSFR} is consistent with the kpc-scale properties of local spiral galaxies and with the Schmidt--Kennicutt relation. 
This suggests that SLSNe can occur in environments that follow the same star-formation law as normal star-forming galaxies.

It is not known whether the environment of SN~2017egm can be regarded as representative of SLSNe. 
The stellar mass is atypical among SLSN hosts, but is comparable to those of SNe Type Ib or Ic (that are not the broad-line type) (e.g., \cite{kell12}). 
The similarity between the environments of Type Ibc SNe and SN~2017egm is also presented in this study for the hydrogen column density. 
This could indicate that the progenitors of SLSNe-I are an extension of Type Ibc SNe. 
Because observations of molecular gas in the environments of SLSNe are very limited, it is important to increase the number of samples to achieve a better understanding of SLSNe.

\begin{ack}
We thank the referee for helpful comments and suggestions. 
We would like to acknowledge Patricia Schady and Janet Ting-Wan Chen for providing their MaNGA data. 
We are grateful to the PDJ collaboration for providing opportunities for fruitful discussions. 
BH is supported by JSPS KAKENHI Grant Number 19K03925. 
This work is supported by the ALMA Japan Research Grant of NAOJ Chile Observatory (NAOJ-ALMA-239). 
This paper makes use of the following ALMA data: ADS/JAO.ALMA\#2018.1.00370.S. 
ALMA is a partnership of ESO (representing its member states), NSF (USA) and NINS (Japan), together with NRC (Canada), MOST and ASIAA (Taiwan), and KASI (Republic of Korea), in cooperation with the Republic of Chile. The Joint ALMA Observatory is operated by ESO, AUI/NRAO and NAOJ. 
The Pan-STARRS1 Surveys (PS1) and the PS1 public science archive have been made possible through contributions by the Institute for Astronomy, the University of Hawaii, the Pan-STARRS Project Office, the Max-Planck Society and its participating institutes, the Max Planck Institute for Astronomy, Heidelberg and the Max Planck Institute for Extraterrestrial Physics, Garching, The Johns Hopkins University, Durham University, the University of Edinburgh, the Queen's University Belfast, the Harvard-Smithsonian Center for Astrophysics, the Las Cumbres Observatory Global Telescope Network Incorporated, the National Central University of Taiwan, the Space Telescope Science Institute, the National Aeronautics and Space Administration under Grant No. NNX08AR22G issued through the Planetary Science Division of the NASA Science Mission Directorate, the National Science Foundation Grant No. AST-1238877, the University of Maryland, Eotvos Lorand University (ELTE), the Los Alamos National Laboratory, and the Gordon and Betty Moore Foundation.
\end{ack}


\end{document}